\documentclass[epj]{svjour}
\usepackage{graphics}
\RequirePackage{amsbsy}
\begin{document}

\title{Can the apparent expansion of the Universe be attributed to an increasing vacuum refractive index ?}

\author{X. Sarazin\inst{1}\thanks{\emph{e-mail:} sarazin@lal.in2p3.fr}, F. Couchot\inst{1}, A. Djannati-Ata\"i\inst{2}, M. Urban\inst{1} 
}

\institute{LAL, IN2P3/CNRS, Universit\'e Paris-Saclay, Orsay, France
          \and
          APC, IN2P3/CNRS, Universit\'{e} Paris Diderot, Observatoire de Paris, CEA/Irfu, Paris, France
}

\date{Received: date / Accepted: date}
% The correct dates will be entered by the editor

\abstract{
H.A. Wilson, then R.H. Dicke, proposed to describe gravitation by a spatial change of the refractive index of the vacuum around a gravitational mass. 
Dicke extended this formalism in order to describe 
the apparent expansion of the Universe by a cosmological time dependence of the global vacuum index.
In this paper, we develop Dicke's formalism.
The metric expansion in standard cosmology (the time-dependent scale factor of the Friedmann-Lema\^itre curved spacetime metric) 
is replaced by a flat and static Euclidean metric with a change with time of the vacuum index. 
We show that a vacuum index increasing with time produces both the cosmological redshift and time dilation, 
and that the predicted evolution of the energy density of the cosmological microwave background is consistent with the standard cosmology. 
We then show that the type Ia supernov\ae \  data, from the joint SDSS-II and SNLS SNe-Ia samples, are well modeled by a vacuum index varying exponentially as $n(t)=exp(t/\tau_0)$, where $\tau_0=8.0^{+0.2}_{-0.8}$~Gyr.
The main consequence of this formalism is that the cosmological redshift  should affect any atom, with a relative decrease of the energy levels of about $-2 \ 10^{-18} \ \mathrm{s}^{-1}$. 
Possibilities for an experimental investigation of this prediction are discussed. 
}

\maketitle

%---------------------------------------------------------------------------------------------------------------------------------------------------------------
\section{Introduction}
%---------------------------------------------------------------------------------------------------------------------------------------------------------------
\label{sec:intro}

In the theory of General Relativity, gravitation is described as the curvature of the spacetime metric. 
The vacuum properties are assumed to be constant and unmodified by the gravitational field. 
Another possible approach, in the weak and static field limit, is to assume a flat Euclidean spacetime metric and to substitute 
the curvature in the vicinity of
a gravitational mass by a spatial change of the vacuum refractive index in its vicinity.
This alternative idea that the action of gravitation on light can be interpreted in terms of an optical medium with a graded vacuum index is 
very old (see for instance the discussion in Eddington's book published in 1920~\cite{Eddington}).
Wilson~\cite{Wilson} was the first to include matter within this paradigm in the early 1920s so that to explain the gravitational 
redshift by the combined change of the vacuum index and the inertial masses.
Following Wilson, Dicke proposed in 1957~\cite{Dicke} a theory aimed to link gravitation to the electromagnetic properties of the vacuum. These ideas have been subsequently 
studied by few other authors (see, e.g.,~\cite{Pauli}, \cite{Landau}, \cite{Felice}, \cite{Evans-1996-2}, \cite{Puthoff}, and~\cite{Unzicker}). 

At the end of his above-mentioned paper, Dicke proposed to extend this formalism by introducing a cosmological time 
dependence of the vacuum index in order to describe the apparent expansion of the Universe in a static Euclidean metric.
The aim of this paper is to investigate and pursue Dicke's model in the modern observational cosmology era. 
This approach is radically different from the standard cosmology, where the cosmological redshift is 
modeled by the expansion of the Friedmann-Lema\^itre metric, through a time-dependent scale factor which is mainly driven 
at the present epoch by a cosmological constant $\Lambda$. 
Here, the cosmological redshift and the time dilation are produced by the combined change with time of the vacuum refractive index and of atomic energy levels. 

It is important to note that the  formalism proposed in this paper is different from the class of models commonly named tired light (TL) theories (see~\cite{tired-light} and references inside),
where, in order to produce the cosmological redshift, an energy loss of photons is assumed but vacuum properties are kept constant.   
As is well known, the TL theories do not produce the cosmological time dilation in SNe-Ia light curves, nor do they conserve the thermal spectrum when redshifted.
Although the speed of light is changing with cosmic time in the framework discussed here, we stress that it should not be confused either with the so-called variable speed of light theories (VSL)~\cite{VSL}. 
These consist mainly of an extension to the standard cosmology with an expanding Friedmann-Lema\^itre metric where a time variation of the speed of light is allowed in addition, as a substitution for the inflation model, in order to solve the problems of horizon and flatness.

The paper is organized as follows. In section~\ref{sec:GR}, we present the Wilson-Dicke formalism of a vacuum refractive index modified by a gravitational mass as an analogy to General Relativity in the static and weak field approximation limit, together with the description of a specific physical mechanism for the gravitational redshift. 
We describe in section~\ref{sec:cosmo} the extension of that model to a vacuum refractive index varying with time and show how it can produce both the observed cosmological redshift and the time stretching of SNe-Ia light curves.
The evolution of the cosmic microwave background is discussed subsequently. 
In section~\ref{sec:sn1afit}, the time dependence of the vacuum refractive index is derived through a fit to the SNe-Ia magnitude-distance diagram.
Finally, the prediction of a cosmological redshift at small scales, down to the atomic scale, is presented in 
section~\ref{sec:redshiftsmallscale} and possible experimental tests are discussed.

%---------------------------------------------------------------------------------------------------------------------------------------------------------------
\section{Vacuum properties varying in space around a static gravitational mass as an analogy to General Relativity }
%---------------------------------------------------------------------------------------------------------------------------------------------------------------
\label{sec:GR}

 Wilson and Dicke proposed to describe the gravitational field as a variation of
the permittivity $\epsilon_0$ and permeability $\mu_0$ of the vacuum.
In the space around a massive body of mass $M$,  at a distance $r$ of this mass, the electromagnetic vacuum properties are replaced by 
\begin{eqnarray}
\nonumber \epsilon_0(r) & = &  n(r) \epsilon_{0,\infty} \\ 
\mu_0(r)       & = & n(r) \mu_{0,\infty}
\end{eqnarray}
where $n(r)$ corresponds to the vacuum refractive index, and $\epsilon_{0,\infty}$ and $\mu_{0,\infty}$ are the vacuum  permittivity and permeability constants in the absence of gravitational mass $M$~\cite{Wilson}\cite{Dicke}.

As shown by Landau and Lifshitz~\cite{Landau}, the Fermat principle for the propagation of light in a stationary gravitational field can be derived from General Relativity as $\delta \int{g_{00}^{-1/2} dl}=0$ where $g_{00}$ is the first component of the metric tensor. Comparing with the conventional Fermat principle $\delta \int{n ds}=0$, the formal analogy between the refraction index and the curved spacetime metric is obtained as 
\begin{eqnarray}
n = g_{00}^{-1/2} dl/ds
\end{eqnarray}
where $dl$ is the local length element and $ds$ the length element observed at infinity, in the absence of gravitational field.
In a static spherical gravitational field, the index varies at first order in $r^{-1}$ as
\begin{eqnarray}
\label{eq:kg}
n(r)=1+\frac{2GM}{rc_{\infty}^2}
\end{eqnarray}
where $c_{\infty}$ is the speed of light in vacuum in the absence of a
gravitational field. 
Thus the speed of light varies in the gravitational field as
$c(r)=c_{\infty}/n(r)$.  In this framework (see \cite{Wilson} and
\cite{Dicke}), a variation in $\epsilon_0$ (and $\mu_0$) affects the
self-energy of a particle and its inertial mass $m$. Therefore
characteristic atomic energies $E_{atom}$ or atomic radii $R_{atom}$ 
are affected by a variation in $\epsilon_0$. 

The Planck constant $\hbar$ and the electrical charge $e$ are assumed to be constant. 
Given the same dependence on index of $\epsilon_0$ and $\mu_0$, the fine structure constant $\alpha=e^2/(4\pi\epsilon_0 \hbar c)$ is hence also constant. 

As can be seen from Maxwell's equations (see the Appendix), an electromagnetic wave propagating in a medium with a spatially variable refractive index, conserves a constant frequency while its wavelength changes as $n^{-1}$.
Therefore, in the Wilson-Dicke model, photons propagate through the graded vacuum refractive index with constant energy but with  momentum varying as $n^{-1}$.

As detailed for instance in ~\cite{Evans-1996-2} or in \cite{Puthoff},  predictions of General Relativity in the weak field approximation such as the deflection of light, the gravitational redshift, or the advance of the perihelion of Mercury, can be calculated in a static and flat metric using the following relations of physical parameters with the vacuum index
\begin{eqnarray}
\label{eq:wdrules}
\nonumber c(r) & =& n^{-1}(r) \times c_{\infty}  \\
\nonumber E_{atom}(r)&=& n^{-1/2}(r) \times  E_{atom,\infty}  \\
\nonumber R_{atom}(r) &=&  n^{-1/2}(r) \times  R_{atom,\infty}  \\
m(r)&=& n^{3/2}(r) \times m_{\infty}  
\end{eqnarray}
where the subscript symbol $\infty$ corresponds to the parameter values in the absence of the gravitational mass $M$. 
We may point out that the {\it measured} value of the speed of light remains constant locally, despite the variation of the vacuum refractive index~\cite{Dicke}. 
Also, as stressed by Dicke, the variation of the rest mass $m$ with index as $ n^{3/2}(r)$ is the only one which satisfies the weak equivalence principle, i.e. the vanishing of gravitational redshift in a free-falling elevator.

To illustrate the Wilson-Dicke model, we describe how the phenomenon of gravitational redshift (or blueshift), as observed for the first time by Pound and Rebka~\cite{PoundRebka}, can be accounted for by a physical mechanism in this formalism.    
An atom at altitude $h$ is in an excited state of transition energy 
\begin{eqnarray}
E_{atom}(R+h)=E_{atom,\infty}/\sqrt{n(R+h)}
\end{eqnarray}
and emits a photon of energy $E_{atom}(R+h)$ ($R$ is the Earth radius). This photon propagates in vacuum to lower altitudes with constant energy. 
At Earth's surface, it reaches a second similar atom of energy 
\begin{eqnarray}
E_{atom}(R)=E_{atom,\infty}/\sqrt{n(R)}
\end{eqnarray}
A Doppler shift corresponding to $\Delta E = E_{atom}(R+h)-E_{atom}(R)$ must be applied to the atoms in order to recover the absorption rate. 
Using the expression of the vacuum refractive index $n$ given in (\ref{eq:kg}) and supposing $h \ll R$, we have at first order in $h/R$
\begin{eqnarray}
\frac{1}{\sqrt{n(R)}}     & = &  1 -  \frac{GM}{R c^2_{\infty}} \\
\frac{1}{\sqrt{n(R+h)}} & = &  1 -  \frac{GM}{R c^2_{\infty}}(1-\frac{h}{R})
\end{eqnarray}
leading to a gravitational blueshift 
\begin{eqnarray}
\Delta E =  \frac{GM}{R c_{\infty}^2} \frac{h}{R} \times E_{atom,\infty}
\end{eqnarray}
in agreement with the measured value and the General Relativity calculation. 

This result can be expressed as well in term of wavelength measurements (spectral shift). Indeed the wavelength of the photon at its emission at altitude $R+h$ is
\begin{eqnarray}
\lambda(R+h)=\frac{h c(R+h)}{E_{atom}(R+h)} = n^{-1/2}(R+h) \times \lambda_{\infty} 
\end{eqnarray}
where $\lambda_{\infty} = h c_{\infty}/E_{atom,\infty}$. 
This photon propagates in vacuum to lower altitudes with constant energy $E_{atom}(R+h)$.  
Arriving at earth's surface, its wavelength $\lambda_{obs}$, measured by the observer, is
\begin{eqnarray}
\lambda_{obs}(R) = \frac{h c(R)}{E_{atom}(R+h)} = \frac{n^{-1}(R)}{n^{-1/2}(R+h)}\times \lambda_{\infty} 
\end{eqnarray}
It is compared to the reference wavelength of the same spectral ray emitted by an atom at the earth's surface
\begin{eqnarray}
\lambda_{ref}(R) = \frac{h c(R)}{E_{atom}(R)} = n^{-1/2}(R) \times \lambda_{\infty} 
\end{eqnarray}
It corresponds to a spectral blueshift $z$ given by
\begin{eqnarray}
z = \frac{\lambda_{obs}-\lambda_{ref}}{\lambda_{ref}} =  - \frac{GM}{R c_{\infty}^2} \frac{h}{R} 
\end{eqnarray}

%---------------------------------------------------------------------------------------------------------------------------------------------------------------
\section{Vacuum properties varying with time as an analogy to the cosmological metric expansion}
%---------------------------------------------------------------------------------------------------------------------------------------------------------------
\label{sec:cosmo}

\subsection{Formalism of a time-dependent vacuum index}

Following the suggestion by Dicke in~\cite{Dicke}, we now extend the Wilson-Dicke formalism to the case of a (cosmic) time dependence of the vacuum refractive index $n(t)$. 
The metric to be used is Euclidean, flat and static, and does not undergo any expansion.
It is defined by the speed of light today, noted $c_0$, at time $t = 0$. 
The cosmological principle (the Universe is spatially homogeneous) is taken to be a fundamental postulate of the theory. This implies that  the time dependence of the vacuum index is spatially uniform, i.e. $n(t)$ is independent of the space coordinates. 

By analogy with the Wilson-Dicke model summarized in (\ref{eq:wdrules}), the physical parameters vary with time $t$ as follows:
\begin{eqnarray}
\label{eq:ctevst}
\nonumber \epsilon_0(t) & = & n(t) \epsilon_{0,0} \\ 
\nonumber \mu_0(t)        & = & n(t) \mu_{0,0}\\
\nonumber c(t)                 & =& n^{-1}(t) \times c_{0}  \\
\nonumber E_{atom}(t)   &=& n^{-1/2}(t) \times  E_{atom,0}  \\
\nonumber R_{atom}(t)   &=& n^{-1/2}(t) \times  R_{atom,0}  \\
m(t)               &=& n^{3/2}(t) \times m_{0}  
\end{eqnarray}
where the subscript symbol $_0$ corresponds to the value of the parameters today, at time $t\!=\!0$, fixing  $n(t \! = \! 0)\!=\!1$. 

As before, the Planck constant $\hbar$ and the electrical charge $e$ are assumed to be constant with time. 
The fine structure constant $\alpha$ is therefore an absolute constant which is in agreement with current experimental tests (see~\cite{alpha} and references therein). 

As noticed by Dicke, there is an important difference between the case of a spatially graded vacuum index, as studied in the previous section, and the case of a spatially homogeneous vacuum index varying with time. 
Indeed, as detailed in the Appendix, Maxwell's equations for an electromagnetic wave propagating through a time-dependent refractive index $n(t)$, 
imply no variation in wavelength but a change in frequency as $n(t)^{-1}$ (this is also discussed in~\cite{Troitskij}). 
We note that this prediction has not yet been confirmed experimentally and is investigated 
in view of controlling light-matter interaction in time-varying structures~\cite{Hayrapetyan}. 
In the present case of a cosmological time-dependent vacuum index, this expectation implies that photons propagate in the Universe with a constant wavelength but with an energy varying as $n(t)^{-1}$. 
In the following, we show that such a time-varying index induces a cosmological redshift 
and that it can also account for the time dilation, as well as for the evolution of the energy density of the cosmological microwave background.

\subsection{Cosmological redshift}
\label{sec:cosmological_redshift}
To illustrate the mechanism producing the cosmological redshift within this formalism, let us imagine a photon emitted by an atom within a galaxy at a measured redshift $z$, at emission time $t_e<0$ and
of atomic transition energy $E_{atom}(t_e)$. The wavelength of the emitted photon is given by 
\begin{eqnarray}
\lambda_{emission} = \frac{hc(t_e)}{E_{atom}(t_e)}
\end{eqnarray}
and remains constant during the propagation. 
The observed wavelength $\lambda_{observed}$ of the detected photon is then equal to its wavelength at emission. 
It is compared to the reference wavelength $\lambda_{reference}$ of the same atomic transition at observation time $t=0$ in the laboratory, given by
\begin{eqnarray}
\lambda_{reference} = \frac{h c_0}{E_{atom,0}} 
\end{eqnarray}
Given the definition of redshift $z$, one has
\begin{eqnarray}
\label{eq:z1}
1+z = \frac{\lambda_{observed}}{\lambda_{reference}} =  \frac{c(t_e)}{c_0} \frac{E_{atom,0}}{E_{atom}(t_e)} 
\end{eqnarray}
Using (\ref{eq:ctevst}), one gets
\begin{eqnarray}
\label{eq:z2}
1+z = n^{-1/2}(t_e)   \ \ t_e<0
\end{eqnarray}
This corresponds to a red-shift if $n^{-1/2}(t_e)>1$, which means a vacuum refractive index increasing with time, i.e. a speed of light decreasing with time.

It is important to note that (\ref{eq:z2}) implies an equivalence between the square root of the vacuum optical index $n^{1/2}$ 
and the time-dependent scale factor $a(t)=1/(1+z)$ of the Friedmann-Lema\^itre curved spacetime metric. 

\subsection{Time dilation}
The observation of a broadening of supernov\ae \ light curves,
proportional to $1+z$, is considered as the direct evidence of a
cosmological time dilation~\cite{sn-time-dilatation}. This effect is
also expected in the formalism discussed here. The
light curves of SNe-Ia result from the formation of $^{56}$Ni, which
decays to $^{56}$Co and then into $^{56}$Fe. The duration of the
process is then driven by the beta-decay half-lives $T_{1/2}(t)$,
which must vary as the inverse of the rest energy of the involved
particles, specifically as $n^{1/2}(t)=1/(z+1)$. 
This might be thought to cause a narrowing proportional to $1+z$,
rather than a time dilation.  However the frequency of the
electromagnetic wave packet, associated to the supernov\ae \ burst,
decreases with time until its detection, as
$n(t)=1/(1+z)^2$. This corresponds to a time broadening 
of the wave packet proportional to $(1+z)^2$, i.e. at twice the rate of the beta-decay half-lives narrowing. 
The overall effect amounts up to a net time broadening of the light curves, proportional to $1+z$, which is equivalent to standard cosmology.

\subsection{The cosmological microwave background}
%---------------------------------------------------------------------------------------------------------------------------------------------------------------
\label{sec:cmb}

In the standard cosmology, the energy $E_{\gamma}$ of the cosmological microwave background radiation (CMB) varies as $a^{-1}(t)=1+z$. The average number of CMB relic photons in an expanding volume is constant, corresponding to a relic photon density $n_\gamma$ varying as $a^{-3}(t)=(1+z)^3$. Therefore the energy density $\mathcal{E}_{\gamma}$ of the CMB radiation varies as $a^{-4}(t)=(1+z)^4$. In the early plasma epoch, photons were in thermal equilibrium with charged particles and the radiation acquired a black body spectrum with an energy density per frequency interval 
\begin{eqnarray}
\label{eq:cmb1}
\mathcal{E}_{\gamma}(\nu) d\nu = \frac{8\pi h}{c^3} \frac{\nu^3 d\nu}{e^{h\nu/k_BT} - 1}
\end{eqnarray}
where $k_B$ is the Boltzmann constant and $T$ is the black body temperature which is related to the radiation energy density by
\begin{eqnarray}
\label{eq:cmb2}
\mathcal{E}_{\gamma} = \frac{\pi^2 k_B^4}{15 \hbar^3 c^3} T^4
\end{eqnarray}
Therefore the temperature $T$ varies as $a^{-1}(t)=(1+z)$.  
After the decoupling, photons no longer interact. Since $\nu \propto (1+z)$ and $T \propto (1+z)$ also, then from equation~(\ref{eq:cmb1}), the black body spectral shape is preserved as the radiation propagates.

In the present framework, the energy of the CMB photons decreases as $n^{-1}(t)$ but the mass energy of the baryons decreases as $n^{-1/2}(t)$. Hence the apparent energy of the photons, relatively to baryons, decreases as $n^{-1/2}(t)=1+z$, as in standard cosmology. 
The average number of relic photons is constant in a volume defined in the static metric. However, in  a volume 
defined with physical rods, the photon density decreases as $n^{-3/2}(t)=(1+z)^3$. Therefore the energy density of the CMB radiation, relatively to baryons, decreases as $(1+z)^4$,  as in standard cosmology. 
If $k_B$ is constant with time (as $\hbar$), then the black body temperature varies as the CMB energy. 
Therefore the apparent (measured) temperature varies as $n^{-1/2}(t)=1+z$, as in standard cosmology, and the  black body spectral shape is also preserved as the radiation propagates. Finally, in a static metric, the ratio of the photons over baryons $n_{\gamma}/n_b$ is constant with time. 
 
Hence, the evolution of the CMB in our model is consistent with the standard cosmology.

%---------------------------------------------------------------------------------------------------------------------------------------------------------------
\section{Constraining the time dependence of the vacuum index with the type Ia supernov\ae}
%---------------------------------------------------------------------------------------------------------------------------------------------------------------
\label{sec:sn1afit}

In  order to constrain the variation time-scale of the vacuum refractive index $n(t)$, we use  
SNe-Ia  as standardizable candles and fit their Hubble diagram. 
Following ~\cite{Betoule-2014}, we use data from the joint analysis of the SDSS-II and SNLS SNe-Ia samples (redshift $0.01<z<1.3$), assuming that 
these  supernov\ae \ exhibit on average the same intrinsic luminosity  at all redshifts, provided that 
corrections are applied for the time stretching of the light curve ($X_1$) as well as for the color at maximum brightness ($C$). 
A linear correction model  leads to the following definition for the standardized distance modulus  $\hat{\mu}$:
\begin{eqnarray}
\label{eq:mu}
\hat{\mu} = m^*_B - (M_B - \alpha X_1 + \beta C) =  5 \log_{10}\left(\frac{d}{10 \mathrm{pc}}\right)
\end{eqnarray}
where $\alpha$ and $\beta$ are the time stretching and color correction factors, 
$d$ is the distance to the source,
$M_B$ is the absolute blue magnitude of SNe-Ia at an a priori chosen distance of 10~pc, 
and $m^*_B$ is the measured peak magnitude in the source rest-frame, taking into account first the apparent broadening of the supernov\ae \ light curves (proportional to $1+z$), and second the apparent photon redshift (also proportional to $1+z$). Same corrections must also be applied in the present formalism, as discussed in the previous sections. 

In our framework, the distance of the source $d$ is defined in the Euclidean metric and is related to the optical path-length by   
\begin{eqnarray}
\label{eq:dL1}
d = \int_{t}^{0} c(t')\,dt' = c_0 \int_{t}^{0} \frac{dt'}{n(t')}
\end{eqnarray}
with $t<0$, the time of photon emission by the supernova, and $t'=0 $, the time of observation. 

We do not have any known theory in hand in order to derive directly the variation of the vacuum refractive index $n(t)$ as a function of the cosmological time $t$. To solve equation~(\ref{eq:dL1}), we assume that the relative variation of the vacuum index is time invariant, at least for the recent period of the Universe (i.e. for redshifts covered by the studied SN-Ia data set) in which the vacuum seems to be dominant. 
This corresponds to a constant relative variation of the vacuum index per absolute time interval, written as
\begin{eqnarray}
\label{eq:dnt}
\frac{dn(t)}{n(t)} = \pm \frac{dt}{\tau_0}
\end{eqnarray}
where $\tau_0$ is assumed to be independent of time at first approximation.
Since the vacuum index is increasing with time, the sign in (\ref{eq:dnt}) is positive. The latter equation leads to an exponential variation of the index 
\begin{eqnarray}
\label{eq:nt}
n(t) = \exp (t/\tau_0) 
\end{eqnarray}
From equation~(\ref{eq:dL1}), one gets
\begin{eqnarray}
d = c_0 \tau_0 (e^{-t/\tau_0}-1) = c_0 \tau_0 (n^{-1}(t)-1), \ \ \ \ t<0
\end{eqnarray}
Remembering that $n^{-1}(t)=(1+z)^2$, it comes
\begin{eqnarray}
\label{eq:dL3}
d = c_0 \tau_0 \left( (1+z)^2-1 \right)
\end{eqnarray}
The estimated standardized distance modulus $\mu$ is then 
\begin{eqnarray}
\label{eq:mup}
\mu(z,\tau_0) = 5 log_{10} \left((1+z)^2-1 \right) + 5 log_{10} \left( \frac{c_0\tau_0}{10 \textrm{pc}} \right)
\end{eqnarray}

This function is fitted to the data by minimizing the $\chi^2$ matrix expression 
\begin{eqnarray}
\label{eq:chi2}
\chi^2 = \left( \hat{\mu}(\alpha,\beta) - \mu(z,\tau_0) \right)^{\dag} \mathrm{C}^{-1} \left(\hat{\mu}(\alpha,\beta) - \mu(z,\tau_0) \right)
\end{eqnarray}
where $\alpha$ and $\beta$ are treated as nuisance parameters and $\mathrm{C}$  is the covariance matrix of the vector of distance modulus estimates $\boldsymbol{\mu}$. It takes into account the measurement errors in the apparent magnitude $m_B^*$ and the stretch and color parameters derived by the analysis of every light curve.

As first pointed in~\cite{Sullivan-2011}, both the absolute magnitude $M_B$ and the 
color correction parameter $\beta$  depend  on host galaxy properties.
Here, we follow~\cite{Betoule-2014} by using a step function to account for this dependency, i.e., 
an offset $\Delta_M$ is added to the absolute magnitude
$M_B$ when the host stellar mass is greater than $10^{10}$ times the
solar mass. 
To keep the simplicity of our approach, rather than fitting $\Delta_M$, we fix it to the best fit value found in~\cite{Betoule-2014}, $\Delta_M=-0.07$.
Moreover, since there is a degeneracy between the two global constants $M_B$ and
$\tau_0$ (which is analog to the degeneracy between $H_0$ and
  $M_B$ in standard cosmology), the value of the absolute magnitude
is fixed to the recently reported value $M_B = -19.25 \pm 0.2$ corresponding to the mean value for 171 nearby SNe-Ia with distance estimates primarily based on Cepheids~\cite{Richardson}.

\begin{figure*} 
\centering
\resizebox{1.\hsize}{!}  {\includegraphics*{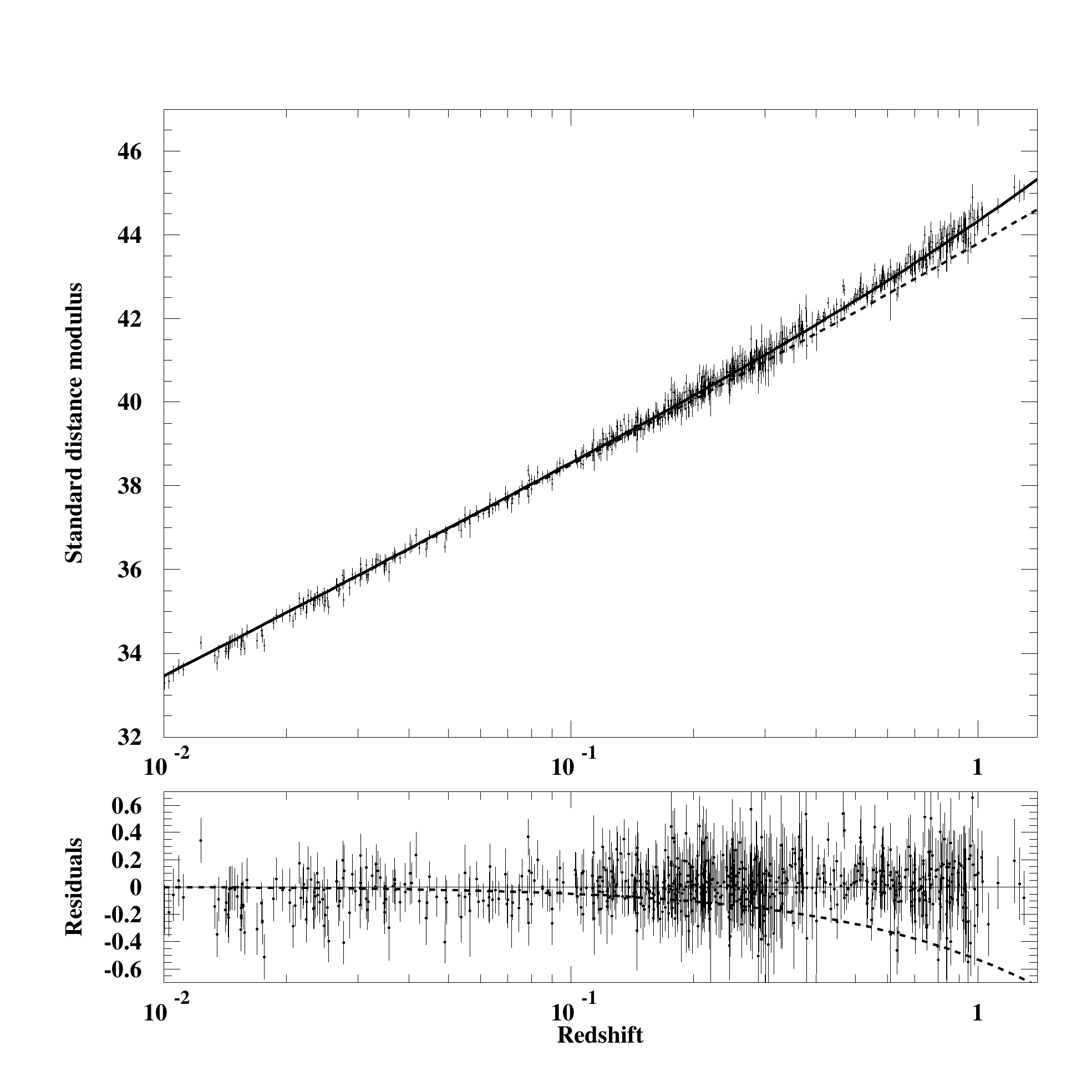}}
\caption{Standard distance modulus $\hat{\mu}$ as a function of redshift $z$ for the combined SDSS-II and SNLS SNe-Ia samples from~\cite{Betoule-2014}. The best fit $\mu$ of the varying vacuum index model is shown as the black line. The solution of the Einstein-de Sitter universe ($\Omega_{M}=1$,$\Omega_{\Lambda}=0$) is presented by the dashed line, for comparison. 
{\it Bottom:} Residuals from the fit as a function of redshift. The dashed line corresponds to the Einstein-de Sitter universe.}
\label{fig:sn-fit-logx}
\end{figure*}

The nuisance parameters $\alpha$ and $\beta$ and the time constant
$\tau_0$ are then the only free parameters.  The result of
the fit is shown in Fig.~\ref{fig:sn-fit-logx}. The fitted
function is in agreement with data with a $\chi^2$ value of 726.3 for
737 degrees of freedom.  
The best fitted values are $\alpha = 0.122 \pm 0.004$ and $\beta=2.612 \pm 0.052$ for the nuisance parameters, and $\tau_0 = 8.0^{+0.2}_{-0.8}$~Gyr for the time constant. 
The error on the latter parameter is dominated by
the systematic uncertainty on the absolute magnitude measurement.  The corresponding relative
variation rate of the vacuum refractive index is $1/\tau_0 = 4 \ 10^{-18} \ \mathrm{s}^{-1}$.  
From (\ref{eq:ctevst}), this implies a relative variation rate of the atomic energy levels of $1/(2\tau_0) = 2 \ 10^{-18} \ \mathrm{s}^{-1}$.

We note that although the hypotheses are very different in the framework proposed here,
the expression of the luminosity distance derived in equation~(\ref{eq:dL3}) is similar to the one obtained in
standard cosmology using the geometry of the Milne universe.
The latter is mathematically equivalent to an empty universe where the metric expansion is only driven by a curvature equal to 1
($\Omega_{M}=\Omega_{\Lambda}=0$, $\Omega_{k}=1$), corresponding to an
expansion scale factor $a(t)$ varying linearly with time. 
The Milne solution can also be obtained assuming a matter-antimatter symmetric cosmology (Dirac-Milne universe), in which antimatter has a negative active gravitational mass~\cite{benoit-levy-chardin}.  
The Milne solution has been lately fitted to the same  supernov\ae \ samples,
leading also to a good agreement with data~\cite{Sarkar-2016}. 
More recently, it has been claimed that the Milne solution is disfavored using a new sample of higher redshift SN-Ia ($1.5 < z <2$)~\cite{Riess-2018}. If this is confirmed, it means that our proposed model of an exponential variation of the vacuum index with a time independent $\tau_0$ constant is only a first order approximation and a characteristic time scale $\tau(t)$ varying with time should be envisaged.

We also note that the exponential variation of the vacuum index $n(t)$  may look similar to the solution of a pure accelerating flat Univers ($\Lambda = 1$) for which the scale factor also varies exponentially ($a(t)=\exp(H_0 t)$). However, in the latter case, the distance varies as $d=c H_0^{-1}(z+z^2)$ which is very different from the solution given in equation~(\ref{eq:dL3}).

%---------------------------------------------------------------------------------------------------------------------------------------------------------------
\section{Cosmological redshift in local bound systems}
%---------------------------------------------------------------------------------------------------------------------------------------------------------------
\label{sec:redshiftsmallscale}

A direct consequence of the proposed model is that the cosmological
redshift should occur not only at cosmological scale but also locally,
since both the vacuum refractive index and the atomic energy levels
are predicted to vary with time at all scales.  

The question of whether the
Hubble apparent expansion of the Universe affects local gravitational systems like
clusters of galaxies, galaxies or planetary systems, has been addressed by many
authors and has received continued studies~\cite{local-expansion}. 
Theoretical considerations conclude that the cosmological expansion of
the Friedmann-Lema\^itre metric must be negligible at the scale of the
galaxy clusters.  However recent measurements of the velocity field of
nearby galaxies in the vicinity of the Local Group have shown a very
linear velocity-distance relation down to about 1~Mpc (i.e.,
approximately the radius of the Local Group), with a local rate of
expansion which coincides with the global Hubble
constant~\cite{local-hubble-measure}.  For very nearby galaxies at
distances smaller than 1~Mpc, measurements are more dispersed because
of undetermined velocity field components.
So we think that any firm conclusion on this matter is excluded for the moment.

It is admitted in standard cosmology that atoms (systems bound by non-gravitational forces) do not expand and that the atomic energy levels do not vary with time. 
In the present formalism, on the contrary, 
the atomic energy levels are predicted to decrease with a relative rate of about $-2 \ 10^{-18} \ \mathrm{s}^{-1}$. 
In order to locally test this shift, let
us consider a thought experiment analog to the Pound$-$Rebka
gravitational redshift experiment. Here one needs 
to measure the redshift of a given atomic transition at two different time periods,
instead of measuring the redshift between two atoms located at two
different gravitational potentials.  Let us study the following setup:
photons emitted by an atomic transition of frequency $f_0$ are
reflected on a mirror located at a distance $D$ and come back to the
initial emitting atoms after a time 
$\Delta t \simeq 2D/c$ (neglecting at first order the variation of the speed of light). 
Comparing the
photon's frequency to the atomic frequency, one expect a net
cosmological redshift with a relative frequency drift of $
\frac{\Delta f}{f_0} = \frac{\Delta t}{2 \tau_0} \approx 2 \ 10^{-18}
\times \Delta t(\mathrm{s})$.  Obviously a very large distance $D$ is
required.

To our knowledge, the longest distance with a single reflection has
been experimentally reached with the satellite Pioneer~10. A radio
signal of frequency $f_0=2.2$~GHz, sent from Earth, reflected on
Pioneer~10, and received back on Earth, allowed for the measurement of
the Doppler velocity of the satellite with a sensitivity (residual
Doppler dispersion after thermal correction) of the order of
10~mHz~\cite{Pioneer10}.  The latest analysed signals have been
obtained when the satellite was at a distance of about 70~A.U.,
equivalent to a round-trip duration of $\Delta t \approx
20$~hours. This corresponds to a predicted cosmological redshift of
about 0.3~mHz,
which is one to two orders of magnitude lower than the residual
Doppler dispersion.  Therefore a positive detection would require the
improvement of the Doppler monitoring accuracy in order to reduce the
residual dispersion.

An alternative way to provide a very large optical path-length is to
use a high-finesse cavity with multiple reflections in the laboratory.
However, because of the decrease of the atomic radii and
consequently of the cavity length as $n^{-1/2}$, the cosmological redshift would be in this case canceled by the
Doppler blueshift induced by the reflection on the mirrors. Alternatively, instead of a rigid cavity, one could think of gravitationally bound mirrors, 
e.g. by using two distant
satellites.  Stability of the atomic frequency must also be controlled
with an ultra high accuracy.  The best accuracy today is achieved in
laboratory with state-of-the-art optical lattice clocks. Frequency
fractional stability of $2 \ 10^{-16}/\sqrt{\tau}$ (where $\tau$ is
the averaging time in seconds) has been recently demonstrated using
$^{87}Sr$ optical lattice clock with a total uncertainty of the clock
of $2 \ 10^{-18}$ in fractional frequency units~\cite{Nicholson}.
Embedding such device in a satellite today is very challenging.
For instance, the frequency stability of the atomic clock of the new project PHARAO~\cite{pharao} (to be installed this year on the 
international space station) is expected to reach few $10^{-14}/\sqrt{\tau}$. 
Although this is orders of magnitude larger than the needed accuracy, further progress  in this type of technology  
could make it possible to test for a local cosmological redshift.

%---------------------------------------------------------------------------------------------------------------------------------------------------------------
\section{Conclusion}
%---------------------------------------------------------------------------------------------------------------------------------------------------------------

As initially proposed by Wilson and Dicke, the curved spacetime in a stationary gravitational field can be equivalently interpreted as being due to a spatial change of the vacuum refractive index and the inertial masses in a Euclidean metric.
Dicke further extended this framework to explain the cosmological redshift, 
assuming a flat and static Euclidean metric but a vacuum index increasing with time. 
The time dependence of the index is postulated to be spatially uniform in order to respect 
the cosmological principle of a spatially homogeneous universe. 
We have investigated Dicke's formalism in the modern observational cosmology era showing
 that it can, remarkably, reproduce not only the cosmological redshift but also the evolution of the CMB energy density and the cosmological time 
dilation of the  supernov\ae \ light curves.  
We have shown in addition that assuming a time-independent variation rate of the vacuum index results in 
a good fit of the SNe-Ia magnitude-distance diagram. Here, an exponential increase of the index $n(t) = \exp(t/\tau_0)$ is obtained with a characteristic time scale of $\tau_0 = 8.0^{+0.2}_{-0.8}$~Gyr. 
Hence the time-dependent scale factor $a(t)$ of the curved spacetime metric in standard cosmology (including the dark energy) can be replaced, at least up to the redshift range covered in the SNe-Ia data used, by a static metric with a vacuum refractive index increasing exponentially with time.

It is important to note that an exponential variation of the vacuum index, if at play up to the highest redshifts,  would correspond 
to the absence of a beginning in the universe evolution, 
and a speed of light infinitely large in an infinitely distant past. 
This would imply in turn that any two given locations in space were causally connected 
in the past, thereby solving the horizon problem without the necessity to recourse to the inflation theory. 
Also, given the ab initio flat space-time used, the observed flatness of the Universe would not require any fine-tuning here.

The present study is far from being complete. No physical mechanism is proposed to account for the variation of the vacuum index, and 
other cosmological probes need to be studied.
Also, possible time dependence of the gravitational constant -- which was one of the prime motivations of Dicke -- could be considered in this framework. 
We hope, however, that this study will stimulate the interest for this unconventional formalism, which could open the path towards alternative 
approaches for solving  current cosmological puzzles such as the dark energy or the horizon problem. 

Finally,  a clear prediction of this work is that the cosmological redshift should affect any atoms -- in deep space, but also  in the laboratory. 
The atomic energy levels are predicted to decrease with time with a relative variation rate of about $-2  \ 10^{-18}$~s$^{-1}$. 
Although very challenging, a single round-trip reflection between two distant satellites could constitute a possible test of this model in future experiments.

%\begin{acknowledgements}
%If you'd like to thank anyone, place your comments here
%and remove the percent signs.
%\end{acknowledgements}

\section*{Appendix: Propagation of an electromagnetic wave in a space or time dependent refractive index}
\label{app}

In this Appendix, we present the properties of electromagnetic plane wave solutions of Maxwell equations for the two cases studied in this article, namely space or time dependent refractive index.
The main results are summarized in Table~\ref{tab:app}.

\subsection*{\it{\textbf{Invariant index}}}
In a linear dielectric and magnetic medium, with permittivity $\epsilon$
and permeability $\mu$, the  Maxwell field equations, expressed with usual SI conventions and notations ($\mathbf{D}=\epsilon\mathbf{E}$ and
$\mathbf{B}=\mu\mathbf{H}$), read
\begin{eqnarray}
\label{eq:Max}
\nonumber \nabla \mathbf{B} & = & 0 \\
\nonumber \nabla \mathbf{D} & = & 0 \\
\nonumber \nabla \times \mathbf{E} & = & -\frac{\partial \mathbf{B}}{\partial t} \\
\nabla \times \mathbf{H} & = & \epsilon\mu\frac{\partial \mathbf{D}}{\partial t}
\end{eqnarray}
The $\mathbf E$ field propagation equation is found from (\ref{eq:Max}) by expressing the double curl of $\mathbf E$ as a function of its derivatives. For constant $\epsilon$ and $\mu$, one gets the simplest propagation equation
\begin{eqnarray}
\label{eq:propn}
\Delta \mathbf{E}  -{\epsilon\mu}\frac{\partial^2 \mathbf{E}}{\partial t^2}=0
\end{eqnarray}
So, when both $\epsilon$ and $\mu$ are real constants, electromagnetic wave
solutions propagate in the medium at a reduced speed $c/n$. The index of refraction $n$ is given by
\begin{eqnarray}
\label{eq:n}
n=\sqrt{\frac{\epsilon\mu}{\epsilon_0\mu_0}}
\end{eqnarray}
A simple solution of (\ref{eq:propn}) is a monochromatic plane wave at frequency $\nu$, with a wavelength $\lambda$, polarized along $x$ and propagating along $z$, defined by
\begin{eqnarray}
\label{eq:solsim}
{E_x}  =   E^0 e^{-i(\omega t - k z)} & {\rm \ ,\ \ } &
{B_y}  =   {n{E_x}}/{c}
\end{eqnarray}
where the angular frequency $\omega=2\pi\nu$ and the wavenumber $k=2\pi/\lambda$ are linked by the dispersion relation
\begin{eqnarray}
\label{eq:dispers}
k=n\omega/c
\end{eqnarray}
The energy density in the wave is evenly shared by $\mathbf{E}$ and $\mathbf{B}$ since
\begin{eqnarray}
\label{eq:epsE2}
n\epsilon_0\mathbf{E}^2 = n\epsilon_0\frac{c^2}{n^2}\mathbf{B}^2 =
\frac{1}{n\mu_0}\mathbf{B}^2
\end{eqnarray}
It is conserved during propagation.
The flux of energy flowing across a $x,y$ plane is given by the Poynting vector
\begin{eqnarray}
\label{eq:Poyncst}
\nonumber\mathbf{S} & = &  \mathbf{E}\times \mathbf{H} \\
S_z & = & \epsilon_0 cE_x^2 
\end{eqnarray}
The flux is also conserved during propagation.

In this article, following Dicke{~\cite{Dicke}}, we take
\begin{eqnarray}
\label{eq:Didi}
\epsilon=n\epsilon_0 {\rm \ \ and \ \ } \mu=n\mu_0
\end{eqnarray}
We first study the case when $n$ varies along the direction of propagation. 
The second case is when $n$ remains uniform in space but varies in time. 

\subsection*{\it{\textbf{Space-dependent index}}}
The case of a space-dependent refractive index $n(z)$ is detailed in Landau{~\cite{Landau-appendix}} for a dielectric and non-magnetic medium. 
When the medium is also magnetic, following the equation~(\ref{eq:Didi}), the propagation equation for the extension of the polarized wave (see (\ref{eq:solsim})) is
\begin{eqnarray}
\label{eq:propn(z)}
\frac{\partial^2 E_x}{\partial z^2}  -\frac{\epsilon\mu}{c^2}\frac{\partial^2 E_x}{\partial t^2}-\frac{1}{\mu}\frac{\partial \mu}{\partial z}\frac{\partial E_x}{\partial z}=0
\end{eqnarray}
By analogy with the solution found in~\cite{Landau-appendix}, the equation~(\ref{eq:propn(z)}) is found to be exactly satisfied by
\begin{eqnarray}
\label{eq:soln(z)}
\nonumber  {E_x}(z,t)  & = & E^0 \exp\Big( -i\omega \big( t - \int_{z_0}^z \frac{n(\zeta)d\zeta}{c}\big) \Big) \\
 {B_y}(z,t)  & = & \frac{n(z){E_x}(z,t)}{c}
%{E_x}(z,t)  = E^0 e^{-i\omega \left( t - \int_{z_0}^z \frac{n(\zeta)d\zeta}{c}\right)}  {\rm\ \ ,\ \ }{B_y}  =  \frac{n(z){E_x}}{c}
\end{eqnarray}
Changing the reference position $z_0$ results in a simple phase shift. In
weak gravitational fields, $n$ varies slowly over a wavelength. So, for $z$ close to $z_0$,(\ref{eq:soln(z)}) resumes to
\begin{eqnarray}
\label{eq:local}
{E_x}(z,t)  \approx E^0 \ \exp\left(-i\omega\left[t  - \frac{n(z_0)}{c}(z-z_0)\right]\right)
%{E_x}(z,t)  \approx E^0 \ e^{-i\omega\left[t  - \frac{n(z_0)}{c}(z-z_0)\right]}
\end{eqnarray}
which is a simple plane wave (see equation~(\ref{eq:solsim})) where the angular frequency is conserved and the wavenumber scales as $n(z)$.

\subsection*{\it{\textbf{Time-dependent index}}}
For the case of a time-dependent refractive index n(t), the propagation equation reads for the $\mathbf{D}$ field
\begin{eqnarray}
\label{eq:propn(t)}
\frac{\partial^2 D_x}{\partial z^2}  - \left(\frac{n}{c}\right)^2  \frac{\partial^2 D_x}{\partial t^2} - \frac{n}{c^2}\frac{\partial n}{\partial t}\frac{\partial D_x}{\partial t}=0
\end{eqnarray}
Its exact solution is
\begin{eqnarray}
\label{eq:solDn(t)}
{D_x}(z,t)  = D^0
\ \exp\left(-i\ k\left[c\int_{t_0}^t\frac{d\tau}{n(\tau)}  - z\right]\right)
\end{eqnarray}
even for quick variations of $n$ compared to the wave period.
This gives for the standard fields
\begin{eqnarray}
\label{eq:soln(t)}
\nonumber  {B_y} (z,t) & = &  B^0\ \exp\left(-i\ k\left[c\int_{t_0}^t\frac{d\tau}{n(\tau)}  - z\right]\right) \\
{E_x}(z,t)  & = & \frac{c B_y(z,t)}{n(t)}
\end{eqnarray}
When $n$ varies slowly with respect to the wave period, (\ref{eq:solDn(t)}) becomes, for $t$ close to $t_0$
\begin{eqnarray}
\label{eq:local}
{D_x}(z,t)  \approx D^0 \ \exp\left(-i\ k\left[c\frac{t-t_0}{n(t_0)}  - z\right]\right)
\end{eqnarray}
which is again a simple plane wave solution (see equation (\ref{eq:solsim})).
Here the wavenumber is conserved, and  the angular frequency scales as $1/n(t)$.

\begin{table}
\centering
\caption{Summary of the main properties of an electromagnetic wave propagating in a space dependent $n(z)$ or a time dependent $n(t)$ refractive index}.
\label{tab:app}
\begin{tabular*}{\columnwidth}{@{\extracolsep{\fill}}lll@{}}
\hline
parameter & $n(z)$ & $n(t)$\\
\hline
E & conserved & $\propto 1/n$\\
B & $\propto n$ & conserved\\
$\omega$ & conserved & $\propto 1/n$\\
$k$ & $\propto n$ & conserved\\
Flux & conserved & $\propto 1/n^2$\\
Energy density & $\propto n$ & $\propto 1/n$\\
\hline
\end{tabular*}
\end{table}

\end{document}